\documentclass[12pt]{article}

\usepackage{amsmath,amsfonts,amssymb}
\usepackage{amsthm}
\usepackage{graphicx}
\usepackage{hyperref}

\newcommand{\slsh} [1] {\not{\hbox{\kern-2pt${#1}$}}}

\newcommand{\beqn}{\begin{eqnarray}}
\newcommand{\eeqn} {\end{eqnarray}}

\newcommand{\beq} {\begin{equation}}
\newcommand{\eeq} {\end{equation}}
\newcommand{\ber}{\begin{eqnarray*}}
\newcommand{\eer} {\end{eqnarray*}}
\newcommand{\bea}{\begin{eqnarray}}
\newcommand{\eea} {\end{eqnarray}}

\newcommand{\Ne}[1]{\ensuremath{\mathcal{N}={#1}}} % Supersymmetry: N = ?
\newcommand{\ls}{\ensuremath{{l_s}}} % String length
\newcommand{\gs}{\ensuremath{{g_s}}} % String coupling
\newcommand{\tym}{{\ensuremath{\theta_{\text{YM}}}}} % \theta_{YM}
\newcommand{\tauym}{{\ensuremath{\tau_{\text{YM}}}}} % \tau_{YM}
\newcommand{\abs}[1]{{\ensuremath{\left\lvert #1 \right\rvert}}} % Absolute value
\newcommand{\vev}[1]{{\ensuremath{\left\langle #1 \right\rangle}}} % Expectation value
\newcommand{\weff}{\ensuremath{W_{\text{eff}}}} % Effective superpotential

\newcommand{\cZZ}[1]{{\ensuremath{\mathbb{C}^3/\mathbb{Z}_{#1}\times\mathbb{Z}_{#1}}}} % C^3 / Z_? x Z_?
\newcommand{\czz}{\cZZ{2}} % C^3 / Z_2 x Z_2

\begin{document}

\thispagestyle{empty}

\begin{flushright}{SWAT/05/437\\
ITP-UU-05/29\\
SPIN-05/23\\
hep-th/0508107
}
\end{flushright}

\vskip 1cm

\centerline{{\Large \bf Predictions for Orientifold Field Theories}}

\vspace{2mm}

\centerline{{\Large \bf  from Type 0' String Theory}}

\vskip 1cm
\centerline{\large Adi Armoni $^{a}$ and  Emiliano Imeroni $^{b}$}

\vskip 0.5cm

\centerline{$^a$\em Department of Physics, University of Wales Swansea,}
\centerline{\it Singleton Park, Swansea, SA2 8PP, UK}
\vspace{2mm}
\centerline{$^b$\em Institute for Theoretical Physics \& Spinoza Institute,}
\centerline{\em Utrecht University, Postbus 80.195, 3508 TD Utrecht, The Netherlands}

\vskip1.3cm

\begin{abstract}

Two predictions about finite-$N$ non-supersymmetric
``orientifold field theories'' are made
by using the dual type 0' string theory on $\czz$ orbifold singularity.
First, the mass ratio between the lowest pseudoscalar and scalar color-singlets
is estimated to be equal to the ratio between
the axial anomaly and the scale anomaly at strong coupling, $M_- / M_+ \sim C_- / C_+$. Second, the ratio between the domain wall tension and the value of the quark condensate is computed.

\end{abstract}

\newpage

\section{Introduction and conclusions}\label{s:intro}

The study of gauge theories at strong coupling is of great importance. 
Our knowledge about the non-perturbative regime of non-supersymmetric
field theories, and QCD in particular, is very limited. In the case
of supersymmetric gauge field theories the situation is drastically
better. In particular, for pure \Ne{1} Super Yang--Mills, certain quantities
(F-terms) are known {\it exactly}. Among them the gluino condensate~\cite{Shifman:1987ia,Davies:1999uw}, the NSVZ beta function~\cite{Novikov:1983uc,Novikov:1985rd} and the tension of the domain walls~\cite{Dvali:1996xe}.

A progress in the understanding of strongly coupled non-supersymmetric gauge theories
was made by arguing for planar equivalence
between \Ne{1} SYM and ``orientifold field theory''~\cite{Armoni:2003gp,Armoni:2004ub}
(see~\cite{Armoni:2004uu} for a review and a comprehensive list of references).
It was shown that at large-$N$ an $SU(N)$ gauge theory with matter
in the antisymmetric two-index representation becomes equivalent to
\Ne{1} SYM in a well defined common bosonic sector. For $SU(3)$ the antisymmetric 
and the antifundamental representations  coincide, hence planar equivalence
can be used to calculate non-perturbative quantities in one-flavor QCD~\cite{Armoni:2003fb}. The common sector of \Ne{1} SYM and the orientifold field theory
includes all Green's functions with gluons as external legs and the
quark condensate.

Planar equivalence already led to a couple of applications, among them
the calculation of the quark condensate in one flavor QCD~\cite{Armoni:2003yv},
an estimate of the lowest scalar and pseudoscalar mass ratio~\cite{Sannino:2003xe}, a new framework of technicolor~\cite{Sannino:2005dy} and a proposal for
a field theory/string theory duality in a non-supersymmetric setup~\cite{DiVecchia:2004ev,DiVecchia:2005vm} (see~\cite{Angelantonj:1999qg} for earlier related work). Orientifold field theories on compact manifolds were discussed
recently in \cite{Barbon:2005zj}.

The string theory description of the orientifold field theory proposed in~\cite{DiVecchia:2004ev,DiVecchia:2005vm} is realized in the framework of the type 0' string, an open 
descendant of the type 0 string~\cite{Sagnotti:1995ga,Sagnotti:1996qj} which includes the same bosonic degrees of freedom and interactions as the type IIB string. This is in agreement with gauge/gravity duality, since the bosonic glueball spectra of \Ne{1} SYM and the orientifold field theory coincide at large-$N$.

In this short paper we wish to use this recently proposed string dual
of orientifold field theory to discuss two pieces of information about the gauge theory.
The first is an estimate of the mass ratio of the lightest glueballs
(similar to the one obtained by Sannino and Shifman \cite{Sannino:2003xe} via
an effective action approach). The second is a prediction for
the domain wall tension to quark condensate ratio. Both predictions
are made for the {\it finite}-$N$ theory.

In general, it is very difficult to calculate $1/N$ corrections, as it involves
loop corrections on the string side. However, we will argue that for the quantities under
consideration the dominant $1/N$ effect is mainly controlled by the R-R background flux,
which in presence of the orientifold plane is shifted from $N$ to $N-2$ (or to $N+2$ in the case of an orientifold field theory with a fermion in the symmetric representation).

The paper is organized as follows: in section \eqref{s:string} we describe an
embedding of orientifold field theory in type 0' string theory. In section \eqref{s:mass} we derive an estimate of the mass ratio between the lightest
scalar and pseudoscalar and finally in section \eqref{s:wall} we derive a
prediction of the ratio between the domain wall tension and the quark condensate.  

\section{A stringy realization of orientifold field theories}\label{s:string}

An explicit stringy realization of orientifold field theories was given by Di Vecchia et al. in~\cite{DiVecchia:2004ev}, in the framework of the type 0B string. Let us first recall that the type 0B string is a non-supersymmetric theory whose closed string spectrum contains only bosons coming from the sectors:
\begin{equation}
	(\text{NS}-,\text{NS}-)\oplus(\text{NS}+,\text{NS}+)
	\oplus(\text{R}-,\text{R}-)\oplus(\text{R}+,\text{R}+)\,.
\end{equation}
In particular, the theory contains a tachyon in the $(\text{NS}-,\text{NS}-)$ sector, and the R-R spectrum (and thus the number of D-branes) is doubled with respect to the type IIB case.

We now give a brief review of a procedure which can be used to engineer non-supersymmetric ``daughter'' gauge theories. We follow~\cite{DiVecchia:2004ev,DiVecchia:2005vm}, where more details can be found. The first step is to take an orientifold of type 0B theory by $\Omega' \mathcal{I}_6 (-1)^{F_L}$, where $\Omega'$ is the world-sheet parity operator, $(-1)^{F_L}$ is the left-moving space-time fermion number and $\mathcal{I}_6$ is the inversion of the six space-time coordinates $x^4,\ldots,x^9$. This theory lives in a non-trivial background given by an O3-plane located at the fixed locus $x^4=\ldots=x^9=0$ of the above transformation.%
\footnote{In the following, we will refer to the orientifold theory described in this section as ``type 0' string theory'', even if, strictly speaking, it is obtained from the usual type 0' theory after six T-dualities.}

One can see that the theory obtained through the orientifold procedure has the same bosonic content as type IIB theory, and in particular the tachyon does not survive the projection. Moreover, the study of the open string sector brings to the conclusion that the gauge theory living on $N$ D3-branes in this orientifold has a bosonic spectrum (one gauge boson and six real scalars) that coincides with the one of \Ne{4} Super Yang--Mills. The fermionic sector consists of four Dirac fermions in the two-index symmetric or antisymmetric representation, depending on the choice of orientifold.

Given the similarity of the gauge theory living on D3-branes in this orientifold with the one living on D3-branes in type IIB theory, it is natural to expect that by performing orbifolds of this orientifold one obtains non-supersymmetric theories related to \Ne{2} and \Ne{1} Super Yang--Mills, and this is indeed what happens.

Let us in particular consider a \czz{} orbifold, defined by the action of two $\mathbb{Z}_2$ generators $g_i$:
\begin{equation}\label{orb}
\begin{tabular}{c|ccc}
	& $z_1$ & $z_2$ & $z_3$ \\
	\hline
	$g_1$ & $z_1$ & $-z_2$ & $-z_3$ \\
	$g_2$ & $-z_1$ & $z_2$ & $-z_3$ \\
\end{tabular}
\end{equation}
where $z^i = x^{2i+2} + i x^{2i+3}$, by the identity $e$ and by $g_3=g_1g_2$.

This orbifold can also be described as the (singular) $F(x,y,z,t)=0$ hypersurface in $\mathbb{C}^4$, where:
\begin{equation}\label{singorb}
        F(x,y,z,t) = xyz + t^2\,,
\end{equation}
where the invariant variables in this function are related to the above complex coordinates by:
\begin{equation}
        x = z_1^2\,,\qquad
        y = z_2^2\,,\qquad
        z = z_3^2\,,\qquad
        t = i \,z_1 z_2 z_3\,.
\end{equation}

There are four types of fractional D3-branes in this background, with different charges under the twisted sectors. If we restrict to a single type of branes, it was shown in~\cite{DiVecchia:2004ev} that the theory living at low energies on the world-volume of $N$ fractional D3-branes is a $U(N)$ theory with the same bosonic content of \Ne{1} Super Yang--Mills (the gauge vector) and whose fermionic matter consists of a single Dirac fermion in the two-index antisymmetric (or symmetric) representation of the gauge group. Namely, we have precisely engineered the desired orientifold field theory.

What kind of information about the orientifold field theory can be derived from the type 0' string theory description? In~\cite{DiVecchia:2004ev,DiVecchia:2005vm}, the gauge theory was analyzed by computing the string annulus partition function with the insertion of an external gauge field, in both the open and closed string channels. The computation yielded the values of the coefficients $C_-$ of the chiral anomaly and $C_+$ of the scale anomaly that, we recall, enter the gauge theory through the anomaly equations
\begin{equation}\label{anomalies}
	\partial_\mu J^\mu = \frac{C_-}{16\pi^2}\, F \tilde{F}\,,\qquad
	T_\mu^\mu = - \frac{3\, C_+}{32\pi^2}\, F^2\,,
\end{equation}
$J^\mu$ being the chiral current and $T^{\mu\nu}$ the standard energy-momentum tensor. It turned out that the computation in the open string channel, as expected, correctly reproduced the known gauge theory results in the field theory limit $\ls\to 0$ (here we concentrate on the theory with antisymmetric matter, similar results hold for the one with symmetric matter):
\begin{equation}\label{openanom}
	C_-^{\text{(open)}} = N - 2\,,\qquad
	C_+^{\text{(open)}} = N + \frac{4}{9}\,.
\end{equation}

On the other hand, the computation in the closed string channel, in the supergravity limit%
\footnote{Here and in the following, the term ``supergravity'' will be used to denote the non-supersymmetric low-energy effective theory of the type 0' string, whose matter content coincides with the bosonic part of type IIB supergravity.}
where only the massless closed strings contribute, yielded the following result:
\begin{equation}\label{closedanom}
	C_-^{\text{(sugra)}} = N - 2\,,\qquad
	C_+^{\text{(sugra)}} = N\,,
\end{equation}
We then see that the chiral anomaly is correctly read also from the supergravity description, which we interpret as a manifestation of the Adler--Bardeen--Jackiw non-renormalization theorem. The fact that the scale anomaly does not coincide with the one obtained in the open string channel was traced in~\cite{DiVecchia:2005vm} to the presence of ``threshold corrections''. However, the scale anomaly is expected to receive corrections, and the strong coupling value need not be the same as the weak coupling value. We would therefore like to argue that the relevant calculation at strong coupling, at least when one limits oneself to the analysis of the lowest-lying glueball states, is the one performed in the supergravity (massless closed string) description, whose result is given in~\eqref{closedanom}.

Let us now try and reinterpret the results of the computation of~\cite{DiVecchia:2004ev} in terms of the geometry generated by the stack of fractional D3-branes (at the orientifold plane). A full solution, analogous to the one found in~\cite{Bertolini:2001gg} for the case of type IIB supergravity, is not available, but we know from the usual dictionary of the gauge/gravity correspondence that the results~\eqref{closedanom} of the closed string computation can be re-expressed in terms of fluxes of the supergravity fields. In particular, comparison with the supersymmetric case studied in~\cite{Bertolini:2001gg,Imeroni:2003cw,Imeroni:2003jk} makes it clear that the geometry generated by the branes will have non-trivial fluxes of the three-form $G_3 = d C_2 + (C_0 + i e^{-\phi})\, d B_2$ along the cycles $A_i$ and $B_i$, $i=1,2,3$ (which are respectively compact and non-compact), which form a standard basis of orthogonal three-cycles on the the Calabi--Yau orbifold geometry:
\begin{equation}\label{gaugegravity}
	\frac{1}{8\pi^2\gs\ls^2} \int_{A_i} G_3 = - (N - 2)\,,\qquad\qquad
	\frac{1}{8\pi^2\gs\ls^2} \int_{B_i} G_3 = - \frac{N}{2\pi i} \ln\frac{r_c}{r_0}\,,
\end{equation}
where the cut-off $r_c$ along $\abs{z_i}$ was introduced due to the non-compactness of the cycles $B_i$ (and a lower scale $r_0$ is further introduced as a short-distance cut-off, since supergravity solutions such as the one in~\cite{Bertolini:2001gg} can usually be shown to be singular at short distances). Notice that the shift by 2 in the flux of $G_3$ through the $A_i$ cycles with respect to the supersymmetric type IIB case is a clear sign of the presence of the O3-plane in the geometry.

The formulas~\eqref{closedanom} and~\eqref{gaugegravity} are the main information that one can extract from the closed string (supergravity) description of the type 0B brane system, and we will see in the following sections how to use them in order to make two predictions about the finite-$N$ behavior of non-supersymmetric $U(N)$ orientifold field theories.

\section{First prediction: Mass ratios}\label{s:mass}

At large-$N$ orientifold field theory becomes equivalent to \Ne{1}
SYM in a well defined sector. This equivalence has many consequences,
among them the degeneracy of even and odd parity hadrons. In this section
we wish to estimate the ratio of the lowest scalar and pseudoscalar masses.

Consider the two-point function $\int d^4 x \,\langle F^2 (0) , F^2(x) \rangle
$. At large-$N$ it is saturated by the free propagator of color-singlet scalars~\cite{Novikov:1981xj}:
\begin{equation}\label{scalars}
\int d^4 x \, e^{iqx} \,\langle F^2 (0) , F^2(x) \rangle \big\rvert_{q^2=0}= \sum _ + \frac{f_+ ^2}{q^2-M_ +^2} \big\rvert_{q^2=0}\,,
\end{equation}
where $f_+$ denotes the coupling of the scalars. Note also that the contact term (a constant) was omitted. The reader should read the lhs of \eqref{scalars} as if it actually contains a contact term (see \cite{Novikov:1981xj} and the appendix of \cite{Witten:1979vv} for a discussion). Let us assume that the sum~\eqref{scalars} is dominated by the lowest scalar
(denoted by $M_+$). Then we can simply write
\begin{equation}\label{scalar}
\int d^4 x \,\langle F^2 (0) , F^2(x) \rangle \sim - \frac{f_+ ^2}{M_ +^2} \,. 
\end{equation}
The relation \eqref{scalar} cannot be justified, unless the mass ratio between the lowest scalar and the next massive scalar is small. 
This assumption is common in lattice simulations where correlation functions are approximated by a single exponent. A similar truncation is made in the gauge/gravity correspondence when the whole string tower is truncated and only supergravity modes are kept.

A relation similar to~\eqref{scalars} can be written for the pseudoscalars:  
\begin{equation}\label{pseudoscalars}
\int d^4 x \, e^{iqx} \,\langle F \tilde F (0) , F \tilde F(x) \rangle \big\rvert_{q^2=0}= \sum _ - \frac{f_- ^2}{q^2-M_ -^2} \big\rvert_{q^2=0}\,. 
\end{equation}
A truncation similar to \eqref{scalar} can be made to write
\begin{equation}
\int d^4 x \,\langle F \tilde F (0) , F \tilde F (x) \rangle \sim - \frac{f_- ^2}{M_ -^2}  . \label{pseudoscalar}
\end{equation}

Combining \eqref{scalar} and \eqref{pseudoscalar} we arrive at the ratio
\begin{equation}
\frac{
\int d^4 x \,\langle F \tilde F (0) , F \tilde F (x) \rangle
}{
\int d^4 x \,\langle F^2 (0) , F^2(x) \rangle 
}
= \frac{f_- ^2}{f_+ ^2} \frac{M_ +^2}{M_ -^2} \label{ratio}
\end{equation}
Remarkably, the ratio~\eqref{ratio} is {\it exact} for \Ne{1} SYM. The reason is that for SUSY theories the lhs of \eqref{ratio} is one, since $F^2$ and $F\tilde F$ sit in the same multiplet. Moreover, masses of even parity and odd parity and their couplings, $f_+$ and $f_-$, are degenerate in supersymmetric theories. 

Due to the relation with \Ne{1} SYM, at large-$N$ the ratio~\eqref{ratio} is also exact for large-$N$ orientifold field theories. Our aim now is to estimate the above ratio at finite-$N$.

Let us start with the axial anomaly~\eqref{anomalies}
\begin{equation}\label{axial}
\partial_\mu J^{\mu} = \frac{C_-}{16 \pi ^2}\, F \tilde F 
\end{equation}
with $C_- = N-2$ for the orientifold field theory with the antisymmetric matter
(and $C_-= N+2$ for the theory with the symmetric matter). Using \cite{Witten:1979vv}, we write down the axial anomaly equation \eqref{axial} between the vacuum and the $ | - \rangle $ state (a state with a single pseudoscalar) 
\begin{equation}\label{axial2}
\langle 0 | \partial_\mu J^{\mu} | - \rangle =\langle 0 | \frac{C_-}{16 \pi ^2} F \tilde F | - \rangle 
\end{equation}
The lhs and rhs of \eqref{axial2} are respectively proportional to
$\lambda_- M _- ^2$ and to $f_- C_-$, where $\lambda_-$ is the coupling of the axial current to the 
pseudoscalar. Thus, reabsorbing all numerical constant into $\lambda_-$, we find
\begin{equation}
\lambda _ - M_- ^2 = f_- C_- \, . \label{f-}
\end{equation}
The above considerations can be repeated for the scale anomaly as well,
\begin{equation}
T^\mu _ \mu = -\frac{3C_+}{32 \pi ^2}\, F^2 \, ,
\end{equation} 
yielding
\begin{equation}
\lambda _ + M_+ ^2 = f_+ C_+ \, . \label{f+}
\end{equation}   
Inserting \eqref{f-} and \eqref{f+} inside \eqref{ratio} we find
\begin{equation}
\frac{
\int d^4 x \,\langle F \tilde F (0) , F \tilde F (x) \rangle
}{
\int d^4 x \,\langle F^2 (0) , F^2(x) \rangle 
}
= \frac{\lambda _- ^2}{\lambda _+ ^2}
\frac{C _+ ^2}{C _- ^2}
\frac{M_ -^2}{M_ +^2} \label{ratio2} \, .
\end{equation}

The relation \eqref{ratio2} is exact for \Ne{1} SYM and large-$N$ orientifold
field theories. The lhs of \eqref{ratio2} is one as well as the ratio
between the couplings $\lambda _+, \lambda _-$ the anomaly coefficients
$C_+, C_-$ and the masses $M_-, M_+$. In fact, we could directly start our discussion from the above equation.

How does \eqref{ratio2} changes as we move from infinite to finite $N$? Clearly,
we can write
\begin{equation}
\frac{
\int d^4 x \,\langle F \tilde F (0) , F \tilde F (x) \rangle
}{
\int d^4 x \,\langle F^2 (0) , F^2(x) \rangle 
}
= 1 + {\cal O} ( 1/N ) \,, \qquad
\frac{\lambda _ -}{\lambda _ +}= 1 + {\cal O} ( 1/N ) \label{corrections}.
\end{equation}
We assume however that the finite-$N$ corrections \eqref{corrections} are 
small. Perhaps the best explanation for that is via supergravity. The above
$1/N$ corrections can be interpreted as corrections in the
string coupling \gs{} -- namely as ``dynamical''
corrections. On the other hand the finite-$N$ correction to the axial anomaly
is large, $N \rightarrow N-2$. In supergravity the source of this correction
is the presence of the orientifold plane that carries R-R charge and hence
the total R-R flux of the $N$ D-branes and the orientifold plane is $N-2$.

We can therefore write
\begin{equation}
\frac{M_-}{M_+} \sim \frac{C_-}{C_+} \label{massr} \, .
\end{equation}
Namely, we predict that the mass ratio between the lowest pseudoscalar and the scalar is equal to the ratio between the axial and scale anomaly coefficients.

As discussed in the previous section, the value of the axial anomaly is $C_-= N-2$, while the one loop value of the scale anomaly coefficient is $N+\frac{4}{9}$. This value however, is valid at weak coupling, where perturbation theory can be trusted. We need however to estimate the value of $C_+$ at strong coupling, and as discussed in section~\ref{s:string} we take the result~\eqref{closedanom} coming from supergravity, namely $C_+=N$.

Our conclusion is that the ratio between the lowest pseudoscalar and scalar
masses is estimated to be
\begin{equation}
\frac{M_-}{M_+} \sim \frac{N-2}{N} \, . \label{prediction1}
\end{equation}

The interest in~\eqref{prediction1} also resides in the fact that it can be checked
by lattice simulations.
The success of the estimate \eqref{prediction1} depends on our assumptions
that the color-singlet dynamics can be approximated by supergravity modes
and moreover that \gs{} is much smaller than $2/N$.

It is tempting to speculate that \eqref{prediction1}
holds down to $N=3$. Since for $SU(3)$ the antisymmetric representation is equivalent
to the fundamental representation, the prediction is that in one flavor QCD
the ratio between the $\eta '$ mass and the $\sigma $ mass is $M_{\eta '} / M_{\sigma } \sim 1/3$.

It is interesting to compare our prediction to a similar one which was
made by Sannino and Shifman \cite{Sannino:2003xe} who used an effective action approach and the one-loop beta function coefficient $3(N+\frac{4}{9})$ as input. They obtained
\begin{equation}
\frac{M_-}{M_+} \sim 1 -\frac{22}{9N} + b \, ,  \label{SS}
\end{equation}
where the coefficient $b(1/N)$ corresponds to $1/N$ corrections that may
shift the ratio $\frac{22}{9N}$, similarly to \eqref{corrections}.

\section{Second prediction: Quark condensate and domain walls}\label{s:wall}

In order to make the second prediction, let us step back for a moment and recall that \Ne{1} gauge theories can be realized in the framework of ``geometric transitions''~\cite{Gopakumar:1998ki,Vafa:2000wi,Cachazo:2001jy}, where one engineers them via configurations of D5-branes wrapped on supersymmetric two-cycles of resolved Calabi--Yau manifolds. The resulting geometry then flows to the one of a deformed manifold, where branes are replaced by fluxes.

In this context, the effective superpotential of the gauge theory is given in terms of the fluxes of the geometry by the following expression~\cite{Taylor:1999ii,Vafa:2000wi}:
\begin{equation}\label{Vafa}
	\weff \propto \sum_{i}
		\left[\ \int_{A_i} G_3 \int_{B_i} \Omega
		- \int_{A_i} \Omega \int_{B_i} G_3\ \right]\,.
\end{equation}
where $G_3$ is the complex three-form field strength of type IIB supergravity, $\Omega$ is the holomorphic $(3,0)$-form of the Calabi--Yau manifold, and $A_i$ and $B_i$ (which are respectively compact and non-compact) form a standard basis of orthogonal three-cycles on the manifold.

The periods of $\Omega$ in the case of the $\czz$ orbifold were computed in~\cite{Imeroni:2003cw} by deforming the singular geometry~\eqref{singorb} with the introduction of a constant parameter $\xi$~\cite{Berenstein:2003fx}:
\begin{equation}\label{deformation}
	F_\xi(x,y,z,t) = xyz + t^2 - \xi^2\,.
\end{equation}
In an appropriate normalization, the periods were found to be:
\begin{equation}\label{omegaperiods}
	\int_{A_i} \Omega = \xi\,, \qquad
	\int_{B_i} \Omega = -\frac{1}{2\pi{\rm i}} \frac{\xi}{3} \ln\frac{\xi}{r_c^3}\,.
\end{equation}

Applying the formula~\eqref{Vafa} to the case of $N$ fractional D3-branes at a \czz{} singularity of the type IIB string then yielded the Veneziano-Yankielowicz effective superpotential \cite{Veneziano:1982ah} of the gauge theory,
\begin{equation}\label{vy}
	\weff = NS \left( 1 - \ln\frac{S}{\Lambda^3} \right)\,,
\end{equation}
after an appropriate identification of geometric and gauge theory quantities which in particular identified the deformation parameter $\xi$ with the gaugino condensate $S$ of the \Ne{1} gauge theory, and the cut-off $r_c$ with the subtraction scale $\mu$.

Can we perform a similar analysis in the type 0' string context? Of course, talking about an effective superpotential for a non-supersymmetric theory does not make sense. However, one may still consider Vafa's formula~\eqref{Vafa} as giving a formal functional $\weff$ of a field $S$ (which now has to be interpreted as the \emph{quark condensate}), whose minimization yields information about physical quantities of the gauge theory (see for instance~\cite{Dijkgraaf:2002wr,Sannino:2003xe}). It is this effective Lagrangian functional we are looking for in our geometric construction.

We can therefore use the results~\eqref{gaugegravity} and~\eqref{omegaperiods} inside~\eqref{Vafa}, obtaining:
\begin{equation}\label{W-or}
	\weff = (N-2) \left( S - S \ln\frac{S}{\mu^{3}\ e^{2\pi i \tauym(\mu)/(N-2)}}\right)\,,
\end{equation}
where $\tauym(\mu)$ is (at \tym=0) the running coupling constant (at strong coupling) coming from~\eqref{closedanom}.
The relation \eqref{W-or} is a prediction of the type 0' string for orientifold field theories.

What physical quantities can we extract from~\eqref{W-or}? First, minimization of \weff{} with respect to $S$ yields the value of the quark condensate:
\begin{equation}\label{S}
	\vev{S} = \mu^3\ e^{2\pi i \tauym(\mu)/(N-2)} e^{2\pi{\rm i}k/(N-2)}\,,
\end{equation}
where $k=0,\ldots,N-3$. We therefore see that the expected number of vacua of the gauge theory is correctly reproduced (as it should, since we know that this construction yields the known result for the chiral anomaly).

A second quantity that can be calculated by using \eqref{W-or} is the tension of a domain wall interpolating between two vacua, labeled respectively by $k$ and $k+q$, $T_{\rm DW} = \abs{\Delta \weff}$:
\begin{equation}\label{T}
T_{\rm DW} = 2 \mu^3 \sin \frac{\pi q}{N-2}
\end{equation}

While the above quantities \eqref{S} and \eqref{T} are subject to certain uncertainties due to the choice of the scale $\Lambda $ by string theory (which is also influenced by $1/N$ corrections which we are not able to evaluate from our procedure), the ratio
\begin{equation}\label{TS}
\frac{T_{\rm DW}}{\abs{\vev{S}}} = 2 (N-2) \sin \frac{\pi q}{N-2}
\end{equation}
is an unambiguous prediction of type 0' string theory for the orientifold field theory. In particular, the prediction consists in the value $2 (N-2)$ of the prefactor (which would simply be $2N$ in pure \Ne{1} Super Yang-Mills).

The shift $N \rightarrow N-2$ with respect to the supersymmetric theory seems to suggest that the quark condensate itself is proportional to $(N-2)\Lambda^3_{\text{QCD}}$.
Establishing such a relation would be useful for the calculation
of the quark condensate in one-flavor QCD~\cite{Armoni:2003yv}.\\

\noindent {\bf Acknowledgments:}
We would like to thank M. Bertolini, P. Di Vecchia, A. Lerda, A. Parnachev, M. Shifman, G. Shore and G. Veneziano for useful discussions and correspondence. AA thanks CERN-TH for a warm hospitality. EI also acknowledges feedback received at the third Simons Workshop held at Stony Brook in August 2005. The work of AA is supported by a PPARC advanced fellowship.

\providecommand{\href}[2]{#2}

\begingroup\begin{thebibliography}{10}

\bibitem{Shifman:1987ia}
M.~A. Shifman and A.~I. Vainshtein, ``On gluino condensation in supersymmetric
  gauge theories. $SU(N)$ and $O(N)$ groups,'' {\em Nucl. Phys.} {\bf B296}
  (1988)
445.
%%CITATION = NUPHA,B296,445;%%.

\bibitem{Davies:1999uw}
N.~M. Davies, T.~J. Hollowood, V.~V. Khoze, and M.~P. Mattis, ``Gluino
  condensate and magnetic monopoles in supersymmetric gluodynamics,'' {\em
  Nucl. Phys.} {\bf B559} (1999) 123--142,
\href{http://www.arXiv.org/abs/hep-th/9905015}{{\tt hep-th/9905015}}.
%%CITATION = HEP-TH 9905015;%%.

\bibitem{Novikov:1983uc}
V.~A. Novikov, M.~A. Shifman, A.~I. Vainshtein, and V.~I. Zakharov, ``Exact
  Gell-Mann-Low function of supersymmetric Yang-Mills theories from instanton
  calculus,'' {\em Nucl. Phys.} {\bf B229} (1983)
381.
%%CITATION = NUPHA,B229,381;%%.

\bibitem{Novikov:1985rd}
V.~A. Novikov, M.~A. Shifman, A.~I. Vainshtein, and V.~I. Zakharov, ``Beta
  function in supersymmetric gauge theories: instantons versus traditional
  approach,'' {\em Phys. Lett.} {\bf B166} (1986)
329--333.
%%CITATION = PHLTA,B166,329;%%.

\bibitem{Dvali:1996xe}
G.~R. Dvali and M.~A. Shifman, ``Domain walls in strongly coupled theories,''
  {\em Phys. Lett.} {\bf B396} (1997) 64--69,
\href{http://www.arXiv.org/abs/hep-th/9612128}{{\tt hep-th/9612128}}.
%%CITATION = HEP-TH 9612128;%%.

\bibitem{Armoni:2003gp}
A.~Armoni, M.~Shifman, and G.~Veneziano, ``Exact results in non-supersymmetric
  large $N$ orientifold field theories,'' {\em Nucl. Phys.} {\bf B667} (2003)
  170--182,
\href{http://www.arXiv.org/abs/hep-th/0302163}{{\tt hep-th/0302163}}.
%%CITATION = HEP-TH 0302163;%%.

\bibitem{Armoni:2004ub}
A.~Armoni, M.~Shifman, and G.~Veneziano, ``Refining the proof of planar
  equivalence,'' {\em Phys. Rev.} {\bf D71} (2005) 045015,
\href{http://www.arXiv.org/abs/hep-th/0412203}{{\tt hep-th/0412203}}.
%%CITATION = HEP-TH 0412203;%%.

\bibitem{Armoni:2004uu}
A.~Armoni, M.~Shifman, and G.~Veneziano, ``From super-Yang-Mills theory to QCD:
  Planar equivalence and its implications,''
\href{http://www.arXiv.org/abs/hep-th/0403071}{{\tt hep-th/0403071}}.
%%CITATION = HEP-TH 0403071;%%.

\bibitem{Armoni:2003fb}
A.~Armoni, M.~Shifman, and G.~Veneziano, ``SUSY relics in one-flavor QCD from a
  new $1/N$ expansion,'' {\em Phys. Rev. Lett.} {\bf 91} (2003) 191601,
\href{http://www.arXiv.org/abs/hep-th/0307097}{{\tt hep-th/0307097}}.
%%CITATION = HEP-TH 0307097;%%.

\bibitem{Armoni:2003yv}
A.~Armoni, M.~Shifman, and G.~Veneziano, ``QCD quark condensate from SUSY and
  the orientifold large-$N$ expansion,'' {\em Phys. Lett.} {\bf B579} (2004)
  384--390,
\href{http://www.arXiv.org/abs/hep-th/0309013}{{\tt hep-th/0309013}}.
%%CITATION = HEP-TH 0309013;%%.

\bibitem{Sannino:2003xe}
F.~Sannino and M.~Shifman, ``Effective Lagrangians for orientifold theories,''
  {\em Phys. Rev.} {\bf D69} (2004) 125004,
\href{http://www.arXiv.org/abs/hep-th/0309252}{{\tt hep-th/0309252}}.
%%CITATION = HEP-TH 0309252;%%.

\bibitem{Sannino:2005dy}
F.~Sannino, ``Light composite Higgs: LCH @ LHC,''
\href{http://www.arXiv.org/abs/hep-ph/0506205}{{\tt hep-ph/0506205}}.
%%CITATION = HEP-PH 0506205;%%.

\bibitem{DiVecchia:2004ev}
P.~Di~Vecchia, A.~Liccardo, R.~Marotta, and F.~Pezzella, ``Brane-inspired
  orientifold field theories,'' {\em JHEP} {\bf 09} (2004) 050,
\href{http://www.arXiv.org/abs/hep-th/0407038}{{\tt hep-th/0407038}}.
%%CITATION = HEP-TH 0407038;%%.

\bibitem{DiVecchia:2005vm}
P.~Di~Vecchia, A.~Liccardo, R.~Marotta, and F.~Pezzella, ``On the gauge/gravity
  correspondence and the open/closed string duality,''
\href{http://www.arXiv.org/abs/hep-th/0503156}{{\tt hep-th/0503156}}.
%%CITATION = HEP-TH 0503156;%%.

\bibitem{Angelantonj:1999qg}
C.~Angelantonj and A.~Armoni, ``Non-tachyonic type 0B orientifolds,
  non-supersymmetric gauge theories and cosmological RG flow,'' {\em Nucl.
  Phys.} {\bf B578} (2000) 239--258,
\href{http://www.arXiv.org/abs/hep-th/9912257}{{\tt hep-th/9912257}}.
%%CITATION = HEP-TH 9912257;%%.

\bibitem{Barbon:2005zj}
J.~L.~F. Barbon and C.~Hoyos, ``Small volume expansion of almost supersymmetric
  large $N$ theories,''
\href{http://www.arXiv.org/abs/hep-th/0507267}{{\tt hep-th/0507267}}.
%%CITATION = HEP-TH 0507267;%%.

\bibitem{Sagnotti:1995ga}
A.~Sagnotti, ``Some properties of open string theories,''
\href{http://www.arXiv.org/abs/hep-th/9509080}{{\tt hep-th/9509080}}.
%%CITATION = HEP-TH 9509080;%%.

\bibitem{Sagnotti:1996qj}
A.~Sagnotti, ``Surprises in open-string perturbation theory,'' {\em Nucl. Phys.
  Proc. Suppl.} {\bf 56B} (1997) 332--343,
\href{http://www.arXiv.org/abs/hep-th/9702093}{{\tt hep-th/9702093}}.
%%CITATION = HEP-TH 9702093;%%.

\bibitem{Bertolini:2001gg}
M.~Bertolini, P.~Di~Vecchia, G.~Ferretti, and R.~Marotta, ``Fractional branes
  and $\mathcal{N} = 1$ gauge theories,'' {\em Nucl. Phys.} {\bf B630} (2002)
  222--240,
\href{http://www.arXiv.org/abs/hep-th/0112187}{{\tt hep-th/0112187}}.
%%CITATION = HEP-TH 0112187;%%.

\bibitem{Imeroni:2003cw}
E.~Imeroni and A.~Lerda, ``Non-perturbative gauge superpotentials from
  supergravity,'' {\em JHEP} {\bf 12} (2003) 051,
\href{http://www.arXiv.org/abs/hep-th/0310157}{{\tt hep-th/0310157}}.
%%CITATION = HEP-TH 0310157;%%.

\bibitem{Imeroni:2003jk}
E.~Imeroni, ``The Gauge/String Correspondence Towards Realistic Gauge
  Theories,''
\href{http://www.arXiv.org/abs/hep-th/0312070}{{\tt hep-th/0312070}}.
%%CITATION = HEP-TH 0312070;%%.

\bibitem{Novikov:1981xj}
V.~A. Novikov, M.~A. Shifman, A.~I. Vainshtein, and V.~I. Zakharov, ``Are all
  hadrons alike?,'' {\em Nucl. Phys.} {\bf B191} (1981)
301.
%%CITATION = NUPHA,B191,301;%%.

\bibitem{Witten:1979vv}
E.~Witten, ``Current algebra theorems for the $U(1)$ 'Goldstone boson','' {\em
  Nucl. Phys.} {\bf B156} (1979)
269.
%%CITATION = NUPHA,B156,269;%%.

\bibitem{Gopakumar:1998ki}
R.~Gopakumar and C.~Vafa, ``On the gauge theory/geometry correspondence,'' {\em
  Adv. Theor. Math. Phys.} {\bf 3} (1999) 1415--1443,
\href{http://www.arXiv.org/abs/hep-th/9811131}{{\tt hep-th/9811131}}.
%%CITATION = HEP-TH 9811131;%%.

\bibitem{Vafa:2000wi}
C.~Vafa, ``Superstrings and topological strings at large $N$,'' {\em J. Math.
  Phys.} {\bf 42} (2001) 2798--2817,
\href{http://www.arXiv.org/abs/hep-th/0008142}{{\tt hep-th/0008142}}.
%%CITATION = HEP-TH 0008142;%%.

\bibitem{Cachazo:2001jy}
F.~Cachazo, K.~A. Intriligator, and C.~Vafa, ``A large $N$ duality via a
  geometric transition,'' {\em Nucl. Phys.} {\bf B603} (2001) 3--41,
\href{http://www.arXiv.org/abs/hep-th/0103067}{{\tt hep-th/0103067}}.
%%CITATION = HEP-TH 0103067;%%.

\bibitem{Taylor:1999ii}
T.~R. Taylor and C.~Vafa, ``RR flux on Calabi-Yau and partial supersymmetry
  breaking,'' {\em Phys. Lett.} {\bf B474} (2000) 130--137,
\href{http://www.arXiv.org/abs/hep-th/9912152}{{\tt hep-th/9912152}}.
%%CITATION = HEP-TH 9912152;%%.

\bibitem{Berenstein:2003fx}
D.~Berenstein, ``D-brane realizations of runaway behavior and moduli
  stabilization,''
\href{http://www.arXiv.org/abs/hep-th/0303230}{{\tt hep-th/0303230}}.
%%CITATION = HEP-TH 0303230;%%.

\bibitem{Veneziano:1982ah}
G.~Veneziano and S.~Yankielowicz, ``An effective Lagrangian for the pure
  $\mathcal{N} = 1$ supersymmetric Yang-Mills theory,'' {\em Phys. Lett.} {\bf
  B113} (1982)
231.
%%CITATION = PHLTA,B113,231;%%.

\bibitem{Dijkgraaf:2002wr}
R.~Dijkgraaf, A.~Neitzke, and C.~Vafa, ``Large $N$ strong coupling dynamics in
  non-supersymmetric orbifold field theories,''
\href{http://www.arXiv.org/abs/hep-th/0211194}{{\tt hep-th/0211194}}.
%%CITATION = HEP-TH 0211194;%%.

\end{thebibliography}\endgroup
\end{document}